\documentclass [floatfix, superscriptaddress, reprint, showpacs, aps, pra, longbibliography,  10pt]{revtex4-1} 

\usepackage{amsmath}
\usepackage{hyperref}
\hypersetup{
    breaklinks=true,
}
\usepackage{graphicx}
\usepackage{bbm}
\usepackage{amssymb}
\usepackage[english]{babel}
\usepackage[latin2]{inputenc}
\usepackage{lmodern}
\usepackage[T1]{fontenc}
\usepackage{wasysym}
\usepackage{breakurl} 
\usepackage{color}

\graphicspath{{Figs/}}

\usepackage{amssymb}

\begin{document}

\title{
Restoring Narrow Linewidth to a Gradient-Broadened Magnetic Resonance by Inhomogeneous Dressing
}

\author{Giuseppe Bevilacqua}
\author{Valerio Biancalana} 
\email{valerio.biancalana@unisi.it}
\affiliation{Dept. of Information Engineering and Mathematics - DIISM, University of Siena - Italy}

\author{Yordanka Dancheva}
\author{Antonio Vigilante}
\affiliation{Dept. of Physical Sciences, Earth and Environment - DSFTA, University of Siena - Italy}

\begin{abstract}
  We  study the  possibility of  counteracting the  line-broadening of
  atomic  magnetic resonances  due  to inhomogeneities  of the  static
  magnetic  field by  means of  spatially dependent  magnetic dressing,
  driven by an alternating field  that oscillates much faster than the
  Larmor  precession   frequency.  We  demonstrate that an intrinsic resonance linewidth of 25~Hz that has been broadened up to hundreds Hz by a magnetic field gradient, can be recovered by  the application of an appropriate
  inhomogeneous   dressing  field. 
	The findings of our experiments may have immediate and important implications, because they enable the use of atomic magnetometers as robust, high sensitivity sensors to detect {\it in situ} the signal from ultra-low-field NMR imaging setups.

\end{abstract}

\date{\today}

\maketitle

\section{Introduction}

We propose a method denominated {\it Inhomogeneous Dressing Enhancement of Atomic resonance} (IDEA) aimed at rendering optical atomic magnetometers suitable to work in an inhomogeneous magnetic field, such as those applied in ultra-low-field (ULF) NMR imaging. The method is based on dressing atoms by means of a strong magnetic field that oscillates transversely with respect to the (inhomogeneous) bias field around which they are precessing, at a frequency much larger than the local Larmor frequencies.

The magnetic dressing of precessing spins with a harmonic high frequency  field was the subject of studies in the late Sixties, when a model based was developed  on a quantum mechanical approach \cite{haroche_prl_70}. In the last decades magnetic dressing was studied and applied in a variety of works dealing with exquisite quantum experiments \cite{silveri_rpp_17},  development of atomic clocks \cite{zanon_prl_12},  manipulation and control of Bose-Einstein condensates 
\cite{beaufils_pra_08, zenesini_lp_10}, ultra-cold collisions \cite{tscherbul_pra_10}, etc. Recently, we re-examined this kind of system in the case of an arbitrary periodic dressing \cite{bevilacqua_pra_12}, making use of a perturbative approach based on the Magnus expansion \cite{magnus_cpam_54} of the time-evolution operator.  An interesting application of magnetic dressing was also studied very recently, in an experiment where critical dressing (matching the effective Larmor frequencies of different species) was applied to improve the sensitivity to small frequency shifts between two dressed species \cite{swank_pra_18}.

Magnetic resonance imaging (MRI) at ULF  is an emerging  method that uses high sensitivity detectors to measure the spatially encoded precession of pre-polarized nuclear spin ensembles in a microTesla-  field \cite{zotev_sst_07}.

Much like in conventional (high field) MRI, the spatial resolution can be achieved with parallelized measurements based on both frequency and phase encoding: a static inhomogeneity in the main field modulus causes the nuclear spin to precess at different frequencies dependent upon one co-ordinate (frequency encoding), while different initial conditions --imposed by pulsed gradients applied prior to the data acquisition-- enable phase encoding, which is used to infer information for the two remaining co-ordinates.

Besides the obvious, dramatic reduction of the precession frequency, the ULF regime comes with other features \cite{tayler_rsi_17} making opportune a general revision of the  standardly applied MRI methodologies. In particular, the ULF regime enables the use of different approaches and techniques for spin manipulation.  An important difference is in the fact that in the ULF it is possible to apply a (dressing) magnetic field that is much stronger than the static one and oscillates at a frequency much higher than the precession frequency. We use this peculiarity to develop a method that restores the functionality of an optical atomic magnetometer (OAM) detector and makes it suited for operating in the presence of the field gradient applied for frequency encoding.

As alternative (non-inductive) detectors, sensors with extremely high sensitivity can be selected among superconducting quantum interference devices (SQUIDs) and OAMs. 
These advanced sensors respond adequately to the low frequency signals characterizing the ULF regime, and may achieve sensitivities at fT/$\sqrt \mathrm{Hz}$ level,  rendering them state-of-the-art magnetometric sensors in MRI, as well as in other applications requiring extreme performance.

The feasibility of the ULF-MRI approach has been demonstrated with both this kinds of these non-inductive sensors \cite{zotev_sst_07, savukov_jmr_13}. ULF MRI is compatible with the presence of other delicate instrumentation and the magnetic detectors can be used to record low-frequency magnetic signals originating from  sources other than nuclear spins.  In particular, hybrid instrumentation enabling multimodal MRI and magnetoencephalography measurements has been proposed and implemented \cite{panu_mrm_12}.

Compared to conventional MRI, ULF operation brings some relevant advantages. The ultimate spatial resolution of MRI is determined by the NMR linewidth, which in turn depends on the absolute field inhomogeneity. A modest relative homogeneity at ULF turns out to be excellent on the absolute scale: very narrow NMR lines with high signal-to-noise ratio can be recorded at ULF with apparatuses that are relatively simple from the point of view of field generation \cite{meriles_sc_01, mcdermott_sc_02, matlachov_jmr_04, burghoff_apl_05, tayler_rsi_17}. The encoding gradients for ULF MRI can also be generated by simple and inexpensive coil systems \cite{mcdermott_pnas_04, zotev_ieee_07}. Further important advantages of the ULF regime in MRI include the  minimization of susceptibility artifacts \cite{seung_mrm_04} and the possibility of imaging in the presence of conductive materials \cite{matlachov_jmr_04, moessle_jmr_06}.

The sample-sensor coupling factor is a key feature, as in any NMR setup, and in the case of SQUID detectors the need for a cryostat may pose limitations. The latter issue makes the alternative choice of OAM detection attractive, together with the much lower maintenance costs and the robustness of the OAM setups. 

The OAM detection of NMR signals is based on probing the time evolution of optically pumped atoms that are magnetically coupled to the sample. In contrast to other solutions proposed, making use of flux transformers \cite{savukov_apl_13, savukov_jmr_14}  and remote detection techniques \cite{xu_pnas_06} for {\it ex-situ} measurements, here we consider the case of  atoms  precessing in a static field that is superimposed upon a small term generated by nuclear spins precessing at a much lower  rate. In this kind of in-situ MRI setup with OAM detection, the static field gradient  applied to the sample for frequency encoding would also affect  the  atomic   precession,  with   severe  degradation   of  the OAM performance, unless a gradient discontinuity was introduced between the sample  and  sensor  locations,  with  the  need  for  coil  geometries
to hinder sample-sensor coupling.

We conceive, test and describe  an approach allowing for the recording of narrow
atomic resonances in spite of the presence of significant field inhomogeneity. The IDEA method is based on counteracting  the atomic frequency spread caused by a defined field  gradient by  means  of a
spatially-dependent dressing  of  the atomic  sample. Using  this
scheme  in a  MRI setup,  the  static and  the (alternating)  dressing
fields inhomogeneously  affect both  the nuclear  sample and  the
atomic  sensor.  However,  marked selectivity  occurs,  because  the
effect of the dressing field depends on the gyromagnetic factor, so that the nuclear precession is substantially unaffected.

 The paper is organized as follows: in the first section (``Experimental setup'') we summarize the features of our atomic magnetometer and we discuss the effects of magnetic field inhomogeneities on the evolution of the atomic sample that constitutes the core of its sensor. The second section (``Method'')  presents the effects of a dressing field on the precession of the magnetized atoms, and the possibilities of using inhomogeneous dressing field to counteract the resonance broadening caused by static field gradients. The achievements in restoring the linewidth of the atomic magnetic resonance are presented in the next section (``Results''), which shows how a high-sensitivity operation of the magnetometer is possible in spite of the presence of large static field inhomogeneity. The last section (``Application to MRI'') is devoted to show how the IDEA method renders the magnetometer suited to detect MRI signals, i.e. let the magnetometer operate in a static field affected by the large inhomogeneity needed to extract spatial information from the nuclear spectra. Here, the applicability of the IDEA method to ULF-MRI --at a proof-of-principle level-- is demonstrated with an unidimensional reconstruction of a water sample. A conclusive part summarizes the main achievements and briefly discusses the potentialities of IDEA in ULF-MRI.

\section{Experimental setup} \label{sec:setup}
\begin{figure}[ht]
   \centering
    \includegraphics [angle=270, width= 0.8 \columnwidth] {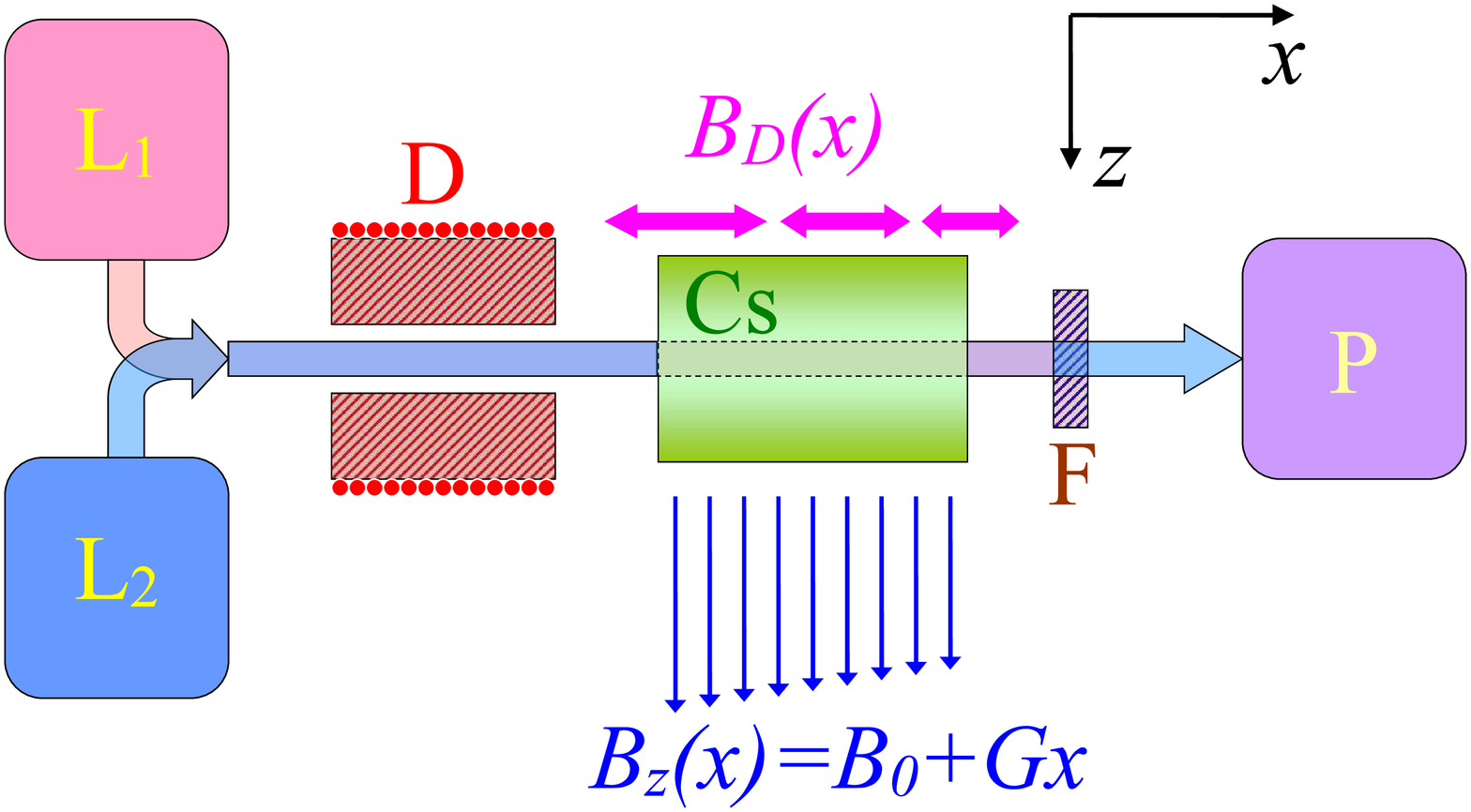}
  \caption{
	Simplified schematics of the magnetometer and of the field arrangement ($L_1$ - pump laser; $L_2$ probe laser; Cs - Cesium cell; F - interference filter stopping the pump radiation; P - balanced polarimeter). The optical axis of the sensor is $x$. 
	A static magnetic field  -- oriented in $z$ direction -- with the main component dependent on the $x$ co-ordinate due to a static quadrupolar term, producing a gradient $G=\partial B_z / \partial x$. 
	The concomitant $\partial B_x / \partial z$ term has no first order effects on the atomic precession  (the presence of a small $x$ component of the field,  amounting at ($\partial B_x / \partial z) \Delta z \ll B_z$ over the beam radius $\Delta z$, has only second-order effects, as also discussed in the Appendix, Eq.\ref{eq:omega2ndorder}). An electromagnetic dipole D oriented along $x$ produces an oscillating (dressing) magnetic field $B_D$ at a  frequency well above the Larmor frequency, oriented along $x$, the strength of which decreases along that direction.
	}
		\label{fig:setup}
\end{figure}

The experimental setup (see Fig.\ref{fig:setup}) is built  around an OAM operating in a Bell \& Bloom configuration, described in detail in Ref.\cite{biancalana_apb_16}. 

Briefly, the OAM uses Cs vapor optically pumped into a
stretched (maximally oriented) state by means of laser
radiation at the milli-Watt level. This pump radiation is
circularly polarized and tuned to the Cs $D_1$ line. The time
evolution of the atomic state is probed by a co-propagating weak (micro-Watt
level) and linearly polarized beam, tuned to the proximity
of the $D_2$ line. A transverse magnetic field B$_0$
causes a precession of the induced magnetization.
The magnetization decay is counteracted via
synchronous optical pumping, which is obtained by modulating
the pump laser wavelength at a frequency $\omega_M/2\pi$, 
which is resonant with the Larmor frequency $\Omega_L/2\pi$.
Scanning $\omega_M$ around $\Omega_L$ makes it possible to characterize the resonance profile. In operative conditions, a resonance width of about $\Gamma=25$~Hz HWHM is measured.

The precession causes a time-dependent Faraday rotation
of the probe radiation. This Faraday rotation is driven to
oscillate at $\omega_M$ (forcing term), to which it responds with
a phase $\varphi$(t) depending on the detuning $\delta=\omega_M-\Omega_L=\omega_M-\gamma B$ ($\gamma$ is the gyromagnetic factor) so as to evolve in accordance with the magnetic field  $B$.
Following interaction with the vapour, the pump radiation is stopped by an interference filter, and the magnetic field and its variation are extracted from the Faraday rotation of the probe beam, as measured by a balanced polarimeter. The sensor, without any passive shielding, operates in a homogeneous B$_0$ field, which is obtained by
partially compensating the environmental field and is oriented along the z axis.  $B_0$ has a typical strength of $4\, \mu$T, giving $T_L= 2\pi / \Omega_L \approx 71\, \mu$s.

The atomic vapor (Cs) is contained in a sealed cell with 23~Torr N$_2$ as a buffer gas, determining a diffusion coefficient D=3.23~cm$^2$/s \cite{franz_pra_74}, which in a precession period causes transverse displacements $\delta = (2D T_L/3)^{1/2}\approx 0.12$~mm.
The laser beam is  $\Phi \approx 1$~cm in diameter, and the condition $\Phi \gg \delta$ enables gradiometric measurements with a base-line of about 5~mm by analyzing the probe spot in two halves \cite{biancalana_apb_16}.

Any increase in the resonance width has detrimental effects on magnetometric sensitivity, rendering it of primary importance to counteract any broadening mechanism. The equation of motion of the magnetization is

\begin{equation}
d\vec M=-\left(\gamma \vec B \times \vec M + D \nabla^2 \vec M +  \Gamma \vec M \right) dt.
\label{eq:diffusion}
\end{equation}

In the case of an inhomogeneous field, the first term in parenthesis is position-dependent and leads to line broadening unless the second (diffusion) term is large enough to make all the atoms behave as if they were precessing around an average field. Indeed, operating with low-pressure cells (i.e. a large D coefficient) may help to counteract gradient-induced resonance broadening, thanks to the so-called motional narrowing (MN) phenomenon \cite{cates_pra_88, pustelny_pra_06}. The MN requires anti-relaxation coating to prevent an increase of the third term in parenthesis due to atom-wall collisions. In the MN regime, the linewidth quadratically depends on the field inhomogeneity, so that MN is only effective with adequately weak gradients. As an example (see Eq.~62 in ref \cite{cates_pra_88}) using a vacuum cell 2~cm in size, MN would maintain a width below $\Gamma$ only for G < 2~nT/cm.
We consider the opposite case, in which the presence of buffer gas makes the diffusion coefficient quite small. With this limit, besides achieving a local response (with the sub-millimetric $\delta$ mentioned above), non-broadened local resonances are obtained, provided that the frequency variation caused by diffusion displacement in a precession period $T_L$ is negligible with respect to the intrinsic width $\Gamma$, i.e. under the condition $G \ll  (\sqrt{3}\Gamma)/(\gamma \delta) \approx 1\,\mu$T/cm, which is much less stringent than that for the MN.

\section{Method}
This section describes the implementation and the principle of operation of the IDEA method. The main  goal of  this work is to counteract the sensitivity degradation of an OAM using a buffered sensor cell in the high-pressure regime, which is placed in a strong linear (quadrupole) magnetic field gradient such as that used for MRI frequency encoding \cite{ansorge_book_16}. 

The method is based on magnetically dressing atoms whose angular momentum is precessing in a static field. This dressing consists in applying a strong time-dependent field that is oriented perpendicularly to the static field and oscillates at a frequency well above the Larmor frequency. Under these conditions, the two momentum components perpendicular to the dressing field evolve in a rather complicated manner, under its direct and time-dependent effect, while the component along the dressing field is not directly coupled and keeps oscillating harmonically, but at an effective Larmor frequency $\Omega_D<\Omega_L$. To this end, a transverse oscillating  field $B_D$, with inhomogeneity along the $x$ direction, is applied by means of a dipole D oriented along $x$ (as represented in Fig.\ref{fig:setup}). Its concomitant gradients produce both transverse ($y$) and longitudinal ($z$) oscillating components in the off-axis interaction region. However, these spurious terms have negligible effects.

Fig.\ref{fig:setup} represents the arrangement for dc (bias, $B_z$)  and ac 
(dressing, $B_D$) field application. The coils for static field  and field gradient
control  are  not  represented, and the  schematics of the optical  part  are also
simplified. $B_z$ is oriented along $z$ and its  gradient $G=\partial B_z/\partial x$ is set by permanent magnets  arranged   in a quadrupolar configuration.  
Thus, the Larmor frequency set by $B_z$ is position-dependent along the optical axis $x$.

The dipolar field $B_D$ is  produced by  a solenoid  wound around  a  ferrite nucleus to generate $B_D$ oriented along $x$,  with  an amplitude  decreasing in that direction. The ferrite nucleus has a hollow-cylinder shape, which permits precise alignment without  hindering the propagation of the laser beams.

The dressing  field $B_D$ oscillates harmonically and has an axial component
\begin{equation}
 B_D(x, t) =\frac{\mu_0}{2\pi}\frac{m(t)}{(x_0+x)^3} = B_{D0}(x) \cos (\omega t),
\end{equation}
where $\mu_0$  is the vacuum permittivity, $m(t)=m_0  \cos (\omega t)$
is  the oscillating  dipole momentum,  $x_0$  is the  position of  the
sensor with  respect to the dipole along  its axis and  $x$ is the
displacement  from  the  sensor   center.  A  time-dependent  current
oscillating at  $\omega  \gg   \gamma  B_z(x)  =
\Omega_L(x)$  induces a magnetic dipole with adjustable intensity. The ferrite and the use of a resonant circuit help to produce a strong oscillating field (several $\mu$T, in our case).

The field $B_D$ alters  the time  evolution  of the
atomic magnetization in  such way as to make its  $x$ component oscillate
harmonically at a dressed  (reduced) angular frequency with respect to
its    unperturbed     precession    around    the     static    field
\cite{bevilacqua_pra_12}:
\begin{equation}
\label{eq:omegaD}
\Omega_{D}(x)=\Omega(x) J_0 \left( \gamma B_{D0}(x) / \omega \right),
\end{equation}
where $J_i(z)$ is the i-th Bessel function of the first kind.

The spatially-dependent  dressing can compensate the $B_z$ inhomogeneity
in a first order approximation. In fact, being $B_z\approx B_0+Gx$,
\begin{equation*}
  \label{eq:eq:omega:svil}
  \begin{split}
  \Omega_D(x) & = \Omega_D(0) + \Omega_D^{\prime}(0) \, x + \frac{1}{2}
  \Omega_D^{\prime \prime}(0) x^2 + O(x^3) = \\
  & = \gamma  B_0 J_0(\alpha) + \gamma \big[ 3 B_0 \alpha
  J_1(\alpha) + G x_0 J_0(\alpha) \big] \frac{x}{x_0} \\
  &\phantom{=} - \frac{3 \alpha \gamma }{2} 
  \big[ (B_0  -2 G x_0) J_1(\alpha)  + 3 \alpha  B_0 J_0(\alpha) \big]
  \left(\frac{x}{x_0}\right)^2 \\
  &\phantom{=} + O( (x/x_0)^3) 
\end{split}
\end{equation*}
where $\alpha = (\mu_0/2\pi) (\gamma m_0)/(\omega x_0^3) $, and the 
condition for compensating the gradient $G$ is thus
\begin{equation}
  \label{eq:def:grad}
  G = -3 \frac{B_0}{x_0} \frac{\alpha J_1(\alpha)}{J_0(\alpha)}, 
\end{equation}
which, for values $\alpha$ of experimental interest (up to $\alpha\approx 1$), results in G values up to $1.7(B_0/x_0)$.

Under compensated conditions (Eq.\ref{eq:def:grad}), in the second order approximation the angular frequency has the expression 
\begin{equation}
  \label{eq:omega:fin}
  \Omega_D(x) \simeq 
  \gamma B_0 \left[ J_0 - \beta\left(\frac{x}{x_0}\right)^2 \right]
\end{equation}
with 
$ \beta=\left(3 \alpha/2 J_0\right)\left( J_0J_1+6 \alpha J_1^2+3 \alpha  J_0^2 \right)$,
and $J_i=J_i(\alpha)$.

It is worth noting that $\beta$ is non-null for any $\alpha$, meaning that a {\it Helmholtz condition} (zeroed quadratic term) would require the application of a secondary --weaker-- oscillating dipole placed at an opportune, smaller distance on the opposite side of the cell. 
 The relevance of the higher order terms neglected in the Taylor expansions reported above may depend on the specification of the dipole (larger second-order terms with the same first-order dressing inhomogeneity would be obtained using a weak, closely located dipole rather than a stronger but more distant one. Further considerations could be made on the importance of quadrupolar and higher order terms in multipolar expansion of the dressing field source. We will provide below (see Fig.\ref{fig:resultsvsBD}) an experimental proof that in our case the neglected, higher-order terms play a role, but do not constitute a substantial problem.

\section{Results}
In this section we present the effects of IDEA on the atomic precession, and we demonstrate its efficiency in recovering narrow atomic magnetic resonance linewidth and the consequent OAM sensitivity. 
\begin{figure}[ht]
   \centering
    \includegraphics [angle=0, width= \columnwidth] {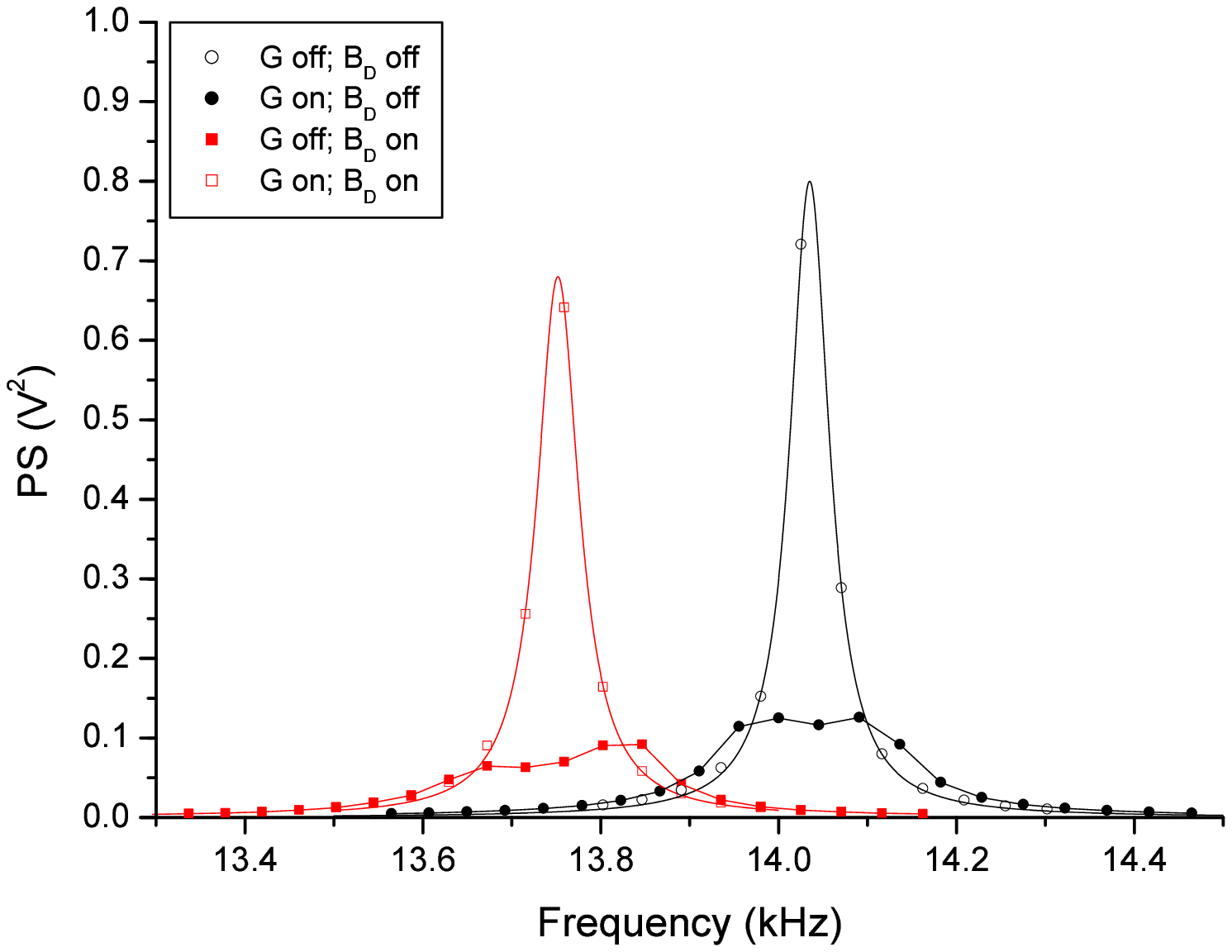} 
  \caption{
  Atomic magnetic resonance (power spectrum PS) under different conditions. The  plot (\Circle) shows the unperturbed resonance; (\CIRCLE) in the presence of a static magnetic field gradient; ({\color{red} $\blacksquare$}) is obtained with no static gradient, but in the presence of a strong, transverse, inhomogeneous field which oscillates much faster than the Larmor precession ($\omega = 2 \pi  32$~kHz). This dressing field (amplitude of about 2.4~$\mu$T) produces a resonance shift that broadens the resonance due to its inhomogeneity (600~nT/cm). With opportune amplitude and frequency values of the dressing field, the two broadening mechanisms compensate each other, and a shifted but narrow resonance can recorded, as shown in the plot ({\color{red} $\square$}). 
    \label{fig:results}}
\end{figure}
We present in  Fig.\ref{fig:results}  a set of spectra 
 obtained under four different conditions, namely in the presence of: a static gradient; the same gradient and appropriate dressing compensation; the same dressing, having removed the static gradient; and with the static homogenous field alone.
The plot (\Circle) in Fig.\ref{fig:results} shows the power spectrum (PS) of the unperturbed resonance in the absence of the gradient and dressing field and under optimal operating conditions. At B$_0$=4$~\mu$T the magnetic resonance amplitude shows a peak at about 14~kHz with a half-width-half-maximum of $\Gamma \approx$25~Hz.   When a quadrupolar magnetic gradient $G=\partial B_z / \partial x=40$~nT/cm is introduced, the resonance gets broader as shown in the plot (\CIRCLE) in Fig.\ref{fig:results}.

In the same figure, the plot ({\color{red} $\blacksquare$}) is obtained 
in the presence of a strong transverse inhomogeneous
field that oscillates much faster than the Larmor precession ($\omega = 2 \pi 32$~kHz) and in the absence of the static gradient.
The dressing field $B_D$ has an amplitude of about 2.4~$\mu$T: it shifts the resonance, and --due to the inhomogeneity-- broadens it, as well. Under appropriate conditions  (Eq.\ref{eq:def:grad}), the two broadening mechanisms compensate each other to the first order, and a shifted but narrow resonance can be recorded, as shown in plot ({\color{red} $\Box$}). The solid lines are  Lorentzian best fits in the cases of narrow resonances (no gradient and dressing-compensated gradient) and eye-guiding interpolations for the two broadened profiles, respectively.

\begin{figure}[ht]
   \centering
    \includegraphics [angle=0, width= \columnwidth] {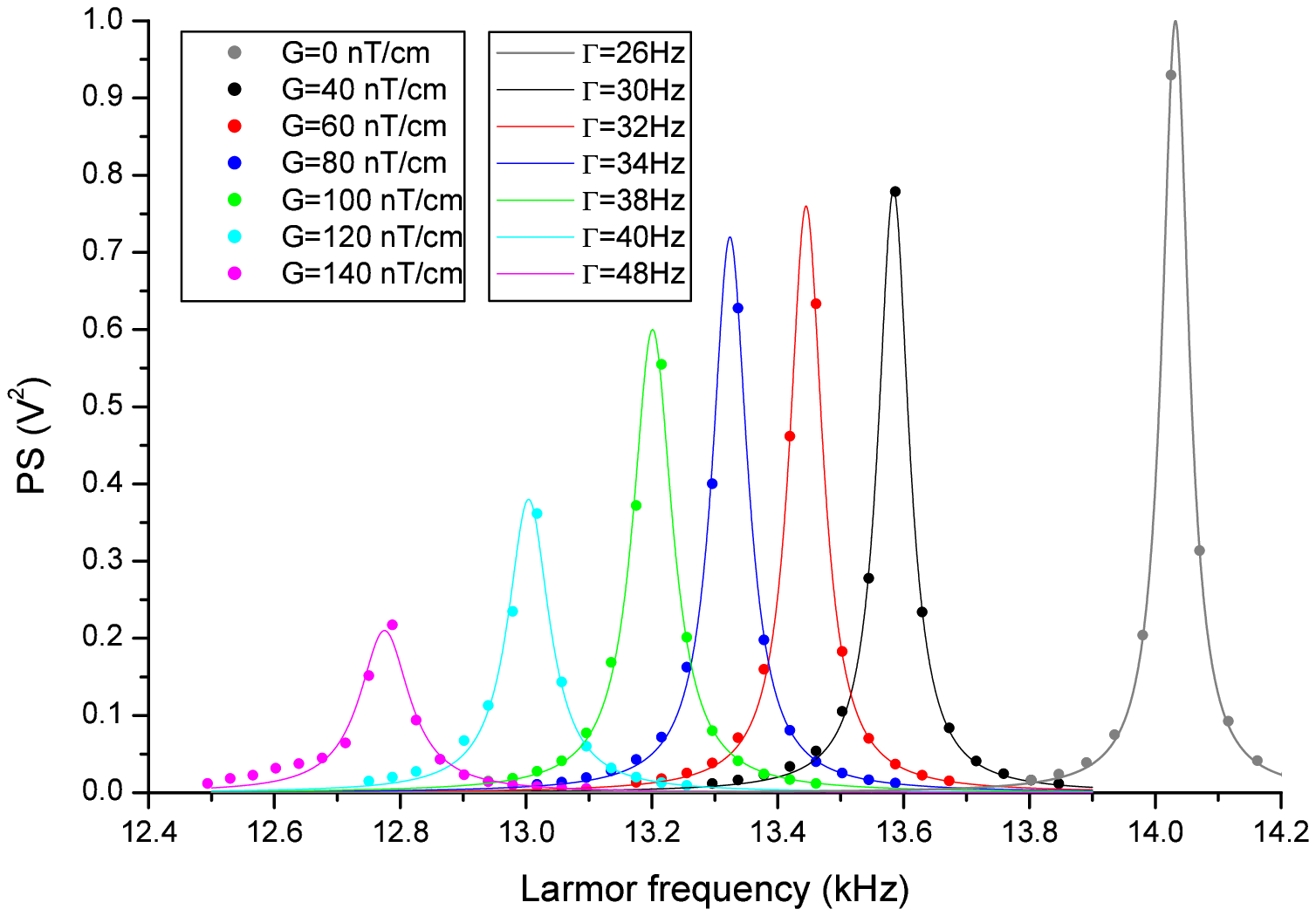} 
  \caption{ Resonance profile (power spectrum PS) under suppressed-broadening conditions, for different values of G. The dots are the amplitudes measured and the lines are best fits targeted to a Lorentzian profile. A progressive shift towards lower frequencies occurs,  consistently with Eq.\ref{eq:omegaD}. Simultaneously, the profile width increases slightly. Above 100~nT/cm, some deviations from the Lorentzian model,  due to the higher order terms (Eq.\ref{eq:omega:fin}), appear as an increase in the low-frequency wing. The leftmost plot (recorded at 140~nT/cm) would appear as a 1~kHz broadened resonance in the absence of the dressing field.
  \label{fig:resultsvsBD}}
\end{figure}

When increasing the values of G, the second order term in Eq.\ref{eq:omega:fin} becomes progressively more important, and the dressing optimization cannot fully restore the original linewidths. Fig.\ref{fig:resultsvsBD} shows the resonance profiles (under the condition Eq.\ref{eq:def:grad}) for different values of G. For the larger values of G (and consequently stronger dressing), the non-linear term of Eq.\ref{eq:omega:fin} causes a deformation of the resonance profile, with the left wing slightly exceeding the Lorentzian values.
However, even at very large G values and correspondingly large dressing field, the line broadening is compensated to an excellent level, meaning that the higher order terms keep playing a substantially negligible role: with G=80~nT/cm a 34~Hz line-width (8~Hz broadening) is achieved, compared with the 560~Hz that would be observed without the  dressing field. 

\section{Application to MRI} 
This section demonstrates the effectiveness of the IDEA method in restoring the OAM performance, showing that --in spite of the gradient applied in an ULF-MRI setup-- the OAM recovers its original sensitivity and can be profitably used to detect the MRI signal. 
We tested the IDEA in a preliminary MRI experiment using remotely polarized protons in tap water, adopting the setup described in Refs.\cite{biancalana_DH_jpcl_17, biancalana_zulfJcoupling_jmr_16}. 
Water protons contained in a 4~ml cartridge (pictured in the upper part of Fig.\ref{fig:image}) were prepolarized in a 1~T field and shuttled to the proximity of the sensor  \cite{biancalana_rsi_14}. The experiment was carried out in unshielded environment, where the environmental magnetic noise was preliminarily reduced using an active stabilization \cite{biancalana_prappl_19} method and then canceled by  measuring differentially on a 5~mm baseline.  An automated system permits long-lasting repeated measurements \cite{biancalana_DH_jpcl_17}, requiring synchronous control of the shuttling system and video-camera checks of its performance; the activation and deactivation of the driving field and field-stabilization system; the application of tipping ($\pi/2$) pulses; DAQ and data elaboration.

\begin{figure}[ht]
   \centering
    \includegraphics [angle=0, width= \columnwidth] {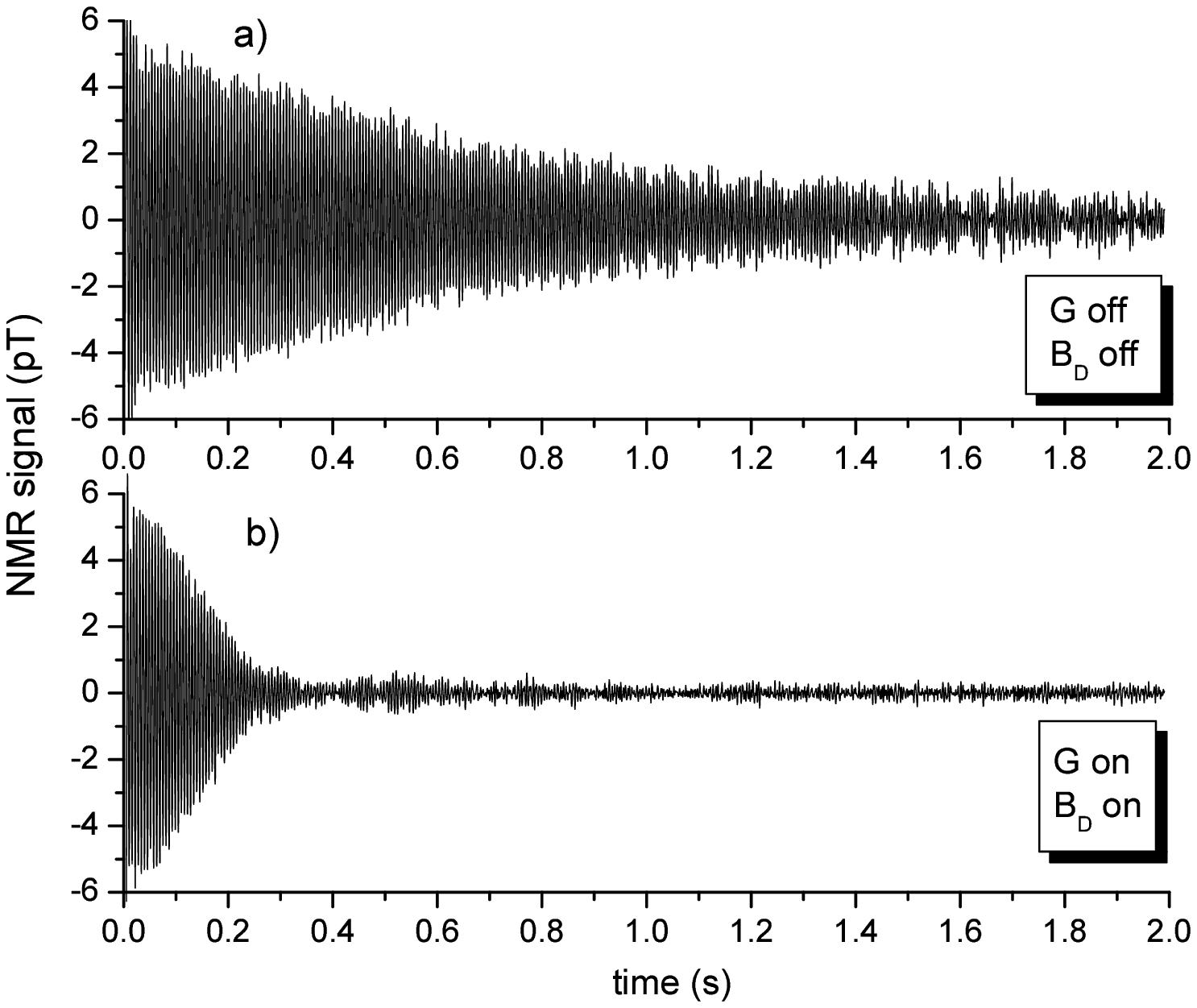} 
  \caption{NMR signal (power spectrum PS) from 4ml H$_2$O sample. Plot a) shows the NMR signal in a homogeneous field with neither  G nor $B_D$ applied. The trace was obtained by averaging over 100 shots.  Plot b) shows the NMR signal recorded  in the presence of $G=50$~nT/cm and $B_D=3\, \mu$T. The trace is obtained by averaging over 400 shots.
  \label{fig:NMRtime}}
\end{figure}

The dressing factor (Eq.\ref{eq:omegaD}) is negligible for the precessing protons due to their much lower gyromagnetic factor, so that in the presence of a static gradient their magnetization precesses at a frequency that depends only on the local static field, as in any frequency-encoded MRI experiment. The time-domain signal recorded appears as shown in Fig.\ref{fig:NMRtime}, with and without the static gradient, respectively. 

\begin{figure}[ht]
   \centering
    \includegraphics [angle=0, width= \columnwidth] {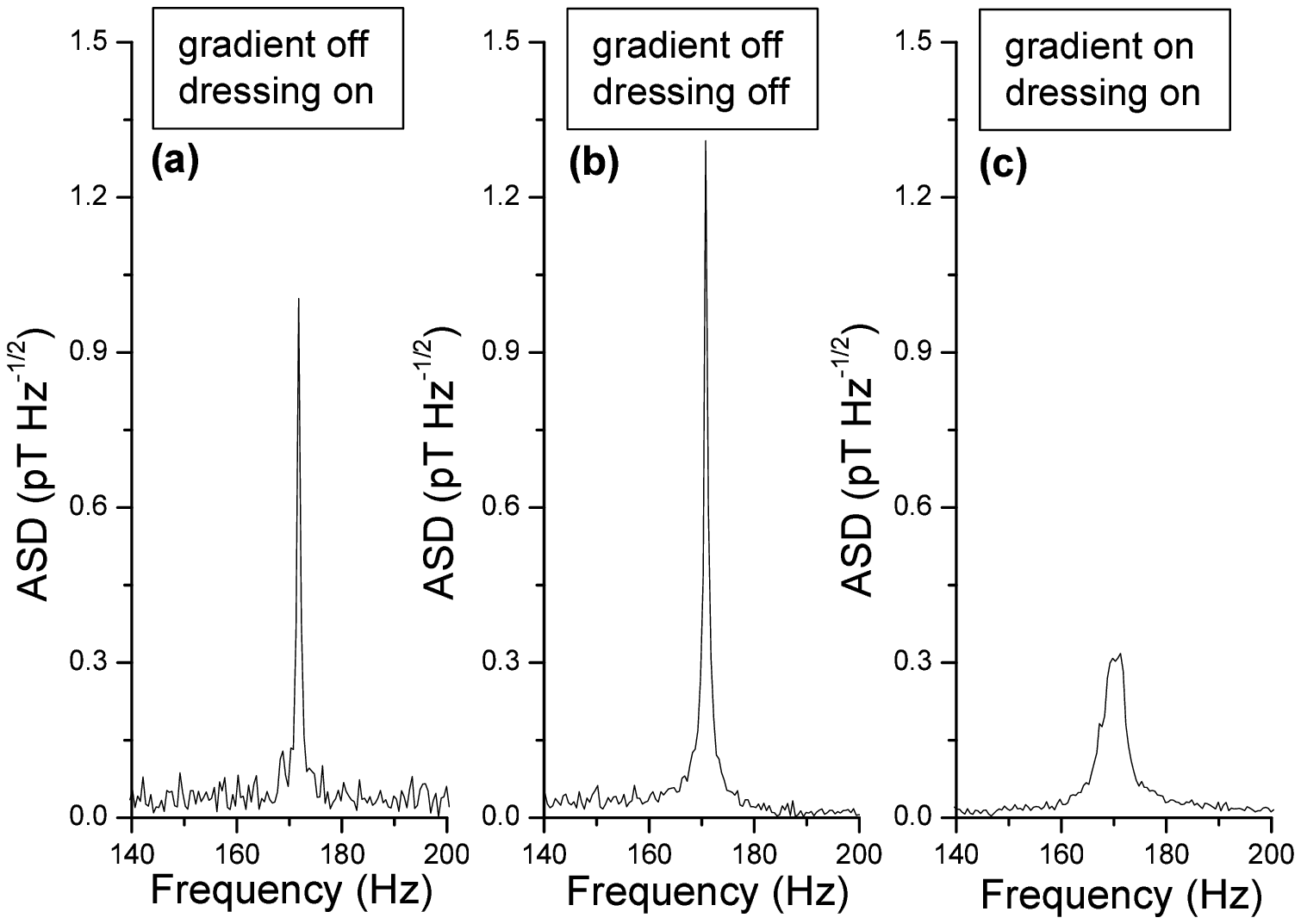} 
  \caption{Tap-water proton NMR amplitude spectral density (ASD) of the signals under different gradient conditions. Plot a) is obtained with $G=0$ and $B_D=3\, \mu$T: a narrow NMR is recorded, with low S/N, because the protons are not affected by $B_D$, while Cs atoms are, so that only a slice of the sensor is effectively pumped. Plot b) is obtained from the data shown  in Fig.\ref{fig:NMRtime} a): $G=0$ and $B_D=0$ produce narrow NMR with good S/N. Plot c) corresponds to the trace shown in Fig.\ref{fig:NMRtime} b): here the static gradient $G$ broadens the NMR spectrum.  The IDEA method allows the same same spectrum to be recorded with a good S/N.
  \label{fig:NMRFFT}}
\end{figure}

\begin{figure}[ht] 
   \centering
    \includegraphics [angle=0, width= \columnwidth] {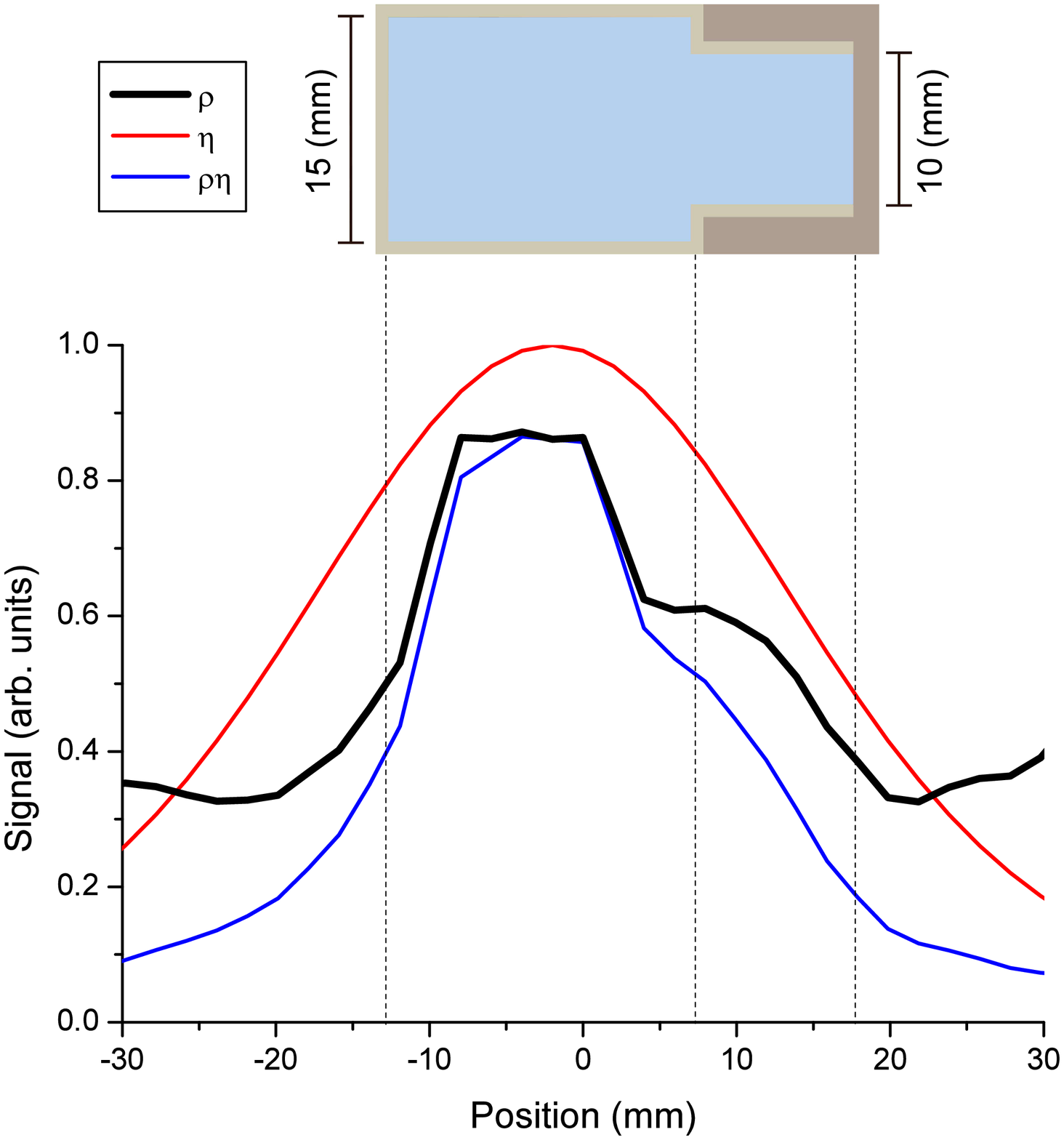} 
  \caption{Plots of the measured $\rho(x) \eta(x)$ (blue), the estimated $\eta(x)$ (red) and the inferred $\rho(x)$ (black), which can be compared with the sample shape represented in the upper part. The data are the same as shown in the plots b) of Figs.\ref{fig:NMRtime} and \ref{fig:NMRFFT}. Some position jitter of the sample occurs and is detected shot-by-shot by the camera. To limit the consequent image blurring, only traces corresponding to sample positioning within a $\pm 3$~mm interval were selected.
  \label{fig:image}}
\end{figure}

Fig.\ref{fig:NMRFFT} shows the effect of the static and dressing field inhomogeneities on the spectra of the proton NMR signal. Plot (a) is obtained in a homogeneous static field while applying a dressing field. The nuclear signal is insensitive to $B_D$ while the dressed Cs atoms have position-dependent resonance, so that only a small fraction (slice) is synchronously pumped and effectively contribute to detect the NMR signal. The resonance recorded has the same width but a worse S/N compared to that resulting for $G=0$ and $B_D=0$ (plot b). The application of a static field $G$ broadens both the atomic and the NMR resonances. However, (plot c) the whole atomic sensor contributes to a broadened NMR signal detection with a good S/N, thanks to IDEA method restoring the atomic resonance linewidth while enabling the registration of position-dependent NMR.

The NMR signal recorded in the presence of the gradient $G$ can be modeled as:
\begin{equation}
S(t)=e^{(-\Gamma_N  - i \omega_0 )t} \int_{-\infty}^{\infty} \eta(x) \rho(x)e^{i \gamma_N G x t}  dx ,
\end{equation}
where $\eta(x)$ represents the detection efficiency  determined by the sample-sensor coupling (see the Appendix for details about the evaluation of $\eta(x)$) and $\rho(x)$ is the proton density in the sample, and $\Gamma_N$ and $\gamma_N$ are the nuclear precession  decay rate and the nuclear gyromagnetic factor, respectively. Following a standard signal elaboration,  after scaling the data by $\exp(\Gamma_N t)$, a Fourier transform is used to reproduce the shape of $\eta(x) \rho(x)$. This is the analysis conducted on the data corresponding to the plots (b) in Figs.\ref{fig:NMRtime} and \ref{fig:NMRFFT} in order to reconstruct the $\eta \rho$ profile shown in Fig.\ref{fig:image}.

\section{Conclusion}
We have proposed, characterized and tested a method (IDEA), based on  magnetically dressing  atomic ground-states, which enables an atomic magnetometer to operate in the presence of a strong field gradient while preserving its sensitivity, thanks to the suppression of gradient-induced resonance broadening.

We found accordance between the theoretical model and the resonance behavior observed. 

We have provided a preliminary demonstration of the applicability of the IDEA method in recording unidimensional NMR images of remotely polarized protons in the ULF regime. 

Our findings suggest that the IDEA method constitutes a  promising tool in ULF MRI using atomic magnetometers. The IDEA method could also be applied in shielded volumes, and in conjunction with phase-encoding techniques, making several kinds of optical magnetometers suitable for use in 3-D MRI apparatuses, in spite of the large gradients that must be applied to achieve fine spatial resolution.

 Considering the application to MRI, the IDEA method makes it possible to operate the OAM in spite of the static field gradient used for the frequency encoding. It is worth noting that the IDEA method  would have no relevance for phase-encoding pulses that should be applied --prior to the measurement-- in the cases of three-dimensional MRI: the non persistent nature of those pulses would make them compatible with the AOM operation.

\section* {Acknowledgements}
The authors are pleased to thank Alessandra Retico for the useful and interesting discussions, Emma Thorley for improving the text, and Ing. Roberto Cecchi for the 3D interactive file provided as a supplemental material (not available with this preprint).

\appendix

\section{Details of MRI geometry and derivation of coupling factor $\eta(x)$} 

This appendix provides a detailed description of the geometry of the MRI detection, and the derivation of the coupling factor $\eta(x)$ used in order to reconstruct the unidimensional sample shape from the magnetometric data obtained with the help of the IDEA method.

\begin{figure}[h!]
   \centering
   \includegraphics [angle=0, width= 0.9 \columnwidth] {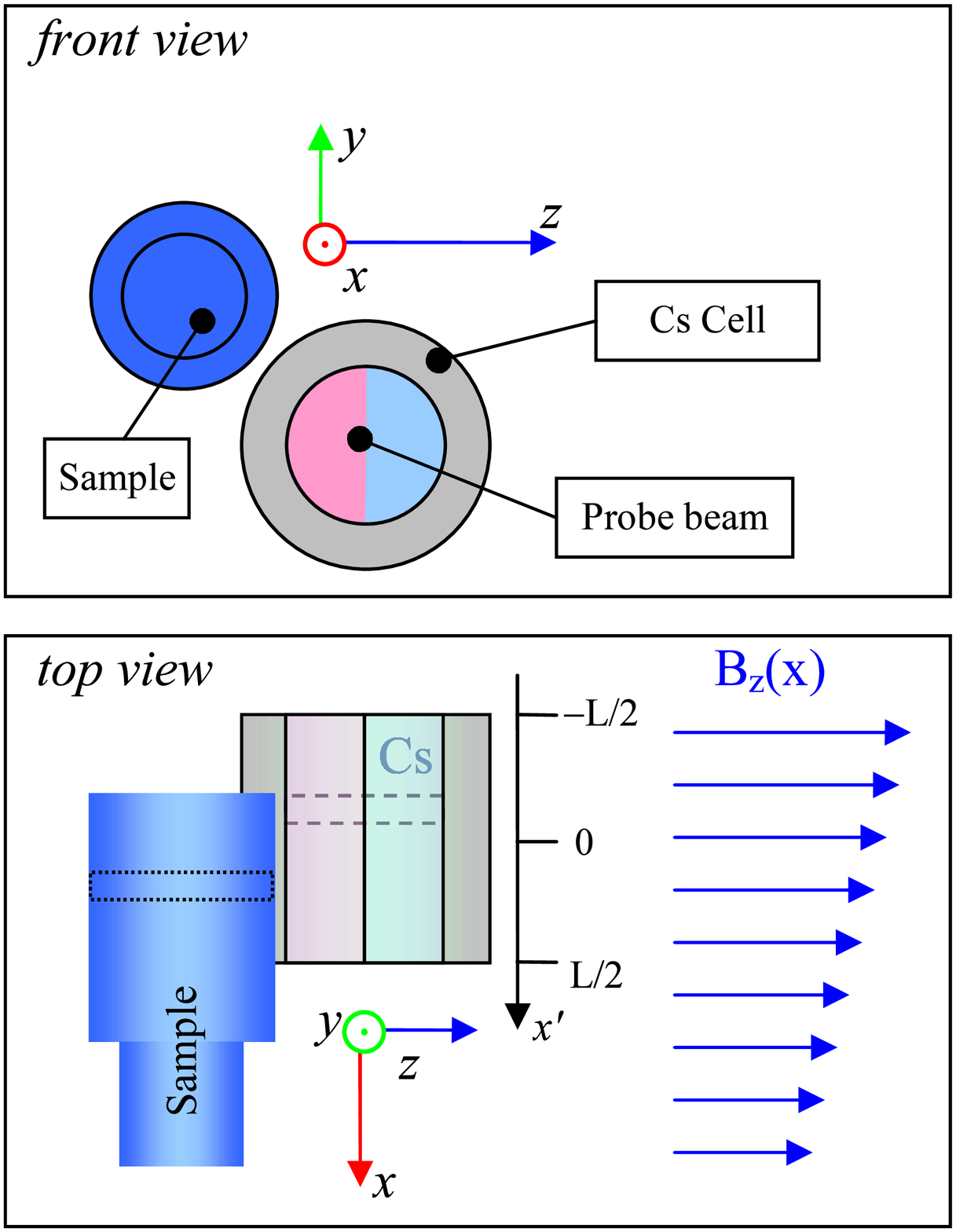}
  \caption{Arrangement of the sensor and sample. Both have cylindrical shapes with axes oriented along $x$. The sensor is a differential one and gradiometrically measures  the $\partial B_z / \partial z$ term of the field generated by the sample. To this aim, the two halves of the probe beam --represented by the two-colored cylinder within the cell-- are analyzed separately. 
  Each slice of the sample (characterized by its $x$ co-ordinate) produces a signal at its specific frequency (MRI frequency encoding). Each slice of the sensor ($x^\prime$ co-ordinate) is differently coupled with various slices of the sample (slices are represented by dashed boxes). The coupling factor $\eta$ is evaluated on the basis of a model in which all the sample slices are modelled as dipolar field sources, and all the sensor slices contribute additively to the polarimetric signal.}
  \label{fig:setupSM}
\end{figure}

Fig.\ref{fig:setupSM} describes the relative positions of the sample and sensor. This figure is complementary to Fig.\ref{fig:setup}: here we highlight the sensor-sample arrangement details and define the variables used to evaluate $\eta(x)$. 

The static magnetic field is oriented along $z$ and varies along the $x$ direction, so that nuclei located in different positions along the axis of the sample precess at different angular frequencies $\omega_N$ (frequency encoding). 

Let $x, y, z$ be co-ordinates of the sample and $x^{\prime}, y^{\prime}, z^{\prime}$ co-ordinates of the sensor.

We model the sample as a linear distribution  of magnetic dipoles along $x$, $\mathbf{m} = (m_x(t), m_y(t), 0)$, with $m_x(t)=m_0 \cos(\omega_N t)$ and $m_y(t)=m_0 \sin(\omega_N t)$, where $\omega_N = \omega_N(x) $: the nuclei precess in the plane $xy$. The magnetic signal is measured with a differential technique to cancel the  environmental disturbances, and to this end the probe beam is analyzed in two beamlets that cross the sensor volume at different $z^{\prime}$. 

Let's derive the response of the sensor to a dipolar field generated by the nuclei at a given $x$. 

The field generated in the position  ${\mathbf r}^\prime$ by a magnetic dipole located  in ${\mathbf r}$ is notoriously:

\begin{equation}
 \mathbf{B} = \frac{\mu_0}{4 \pi}\left[ 3 \frac{\left (\mathbf{m}  \cdot (\mathbf{r}-\mathbf{r^\prime})\right)(\mathbf{r}-\mathbf{r^\prime}) }{|\mathbf{r}-\mathbf{r^\prime}|^5}-\frac{\mathbf{m} }
 {|\mathbf{r}-\mathbf{ r^\prime}|^3}\right].
  \label{eq:Bvect}
\end{equation}

In the presence of a dominant bias field, $\mathbf{ B_0}$, due to the scalar response of our OAM, only the component of $\mathbf{B}$ parallel to $\mathbf {B_0}$ generates a detectable (first order response) signal, because the local Larmor precession of atoms occurs at 
\begin{equation}
\omega=\gamma |B|=\gamma \sqrt{(\mathbf{B_0}+ \mathbf{B})^2} \approx \gamma B_0 \left [1+ \frac{\left (\mathbf{ B_0} \cdot \mathbf{ B} \right )}{B_0^2} \right]. 
\label{eq:omega2ndorder}
\end{equation}
In our case the bias field $\mathbf {B_0}$ is oriented along $z$ so that only the component $B_z$ of the nuclear field is effectively detected.

Thus, in the geometry of our setup ($\mathbf {m}$ in the $xy$ plane), the second term in the squared parenthesis of Eq.\ref{eq:Bvect} does not contribute to the field measured, and the relevant quantity for the local effect of a sample slice is

\begin{equation}
 B_z = \frac{3 \mu_0}{4 \pi} \frac{\left [\mathbf{m} \cdot (\mathbf{r}-\mathbf{r^\prime})\right](z-z^\prime) }{|\mathbf{r}-\mathbf{ r^\prime}|^5}.
 \label{eq:BzsliceVec}
\end{equation}

Neglecting the dimensions along $y$ and $z$ and defining $\Delta y=(y-y^\prime)$,  Eq. \ref{eq:BzsliceVec} turns into:
\begin{equation}
\begin{split}
 B_z & = \frac{3 \mu_0}{4 \pi}  \frac { \left [ m_x(x-x^\prime)+ m_y \Delta y \right ](z-z^\prime )}{[ (x-x^\prime)^2+ \Delta y^2+ (z-z^\prime)^2 ]^{5/2}}=
 \\
  & = f \cos(\omega_N t)+ g \sin(\omega_N t),
 \label{eq:Bzslice2}
 \end{split}
\end{equation}
with (making the dependencies explicit)
\begin{equation*}
 f (x^\prime, z^\prime)= \frac{3 \mu_0}{4 \pi}  \frac {  m_0 (x-x^\prime )(z-z^\prime )}{[ (x-x^\prime)^2+ \Delta y^2+ (z-z^\prime)^2 ]^{5/2}}
 \label{eq:f}
\end{equation*}
and
\begin{equation*}
 g (x^\prime, z^\prime)= \frac{3 \mu_0}{4 \pi}  \frac {   m_0 \Delta y (z-z^\prime )}{[ (x-x^\prime)^2+ \Delta y^2+ (z-z^\prime)^2 ]^{5/2}}.
 \label{eq:Bzsliceg}
\end{equation*}

We assume that the response of the sensor is homogeneous along  $x^\prime$. The effect of $f$ and $g$ on the probe beam is obtained by integrating Eq. \ref{eq:Bzslice2} over the sensor length $L$:
\begin{equation}
\begin{split}
 S(t) & =\cos(\omega_N t)\int_{-L/2}^{L/2}f (x^\prime, z^\prime) dx^\prime +\\
 & +\sin(\omega_N t)\int_{-L/2}^{L/2} g (x^\prime, z^\prime) dx^\prime =\\
 & =F(z^\prime) \cos(\omega_N t)+ G (z^\prime) \sin(\omega_N t),
 \label{eq:integral}
\end{split}
\end{equation}
with a resulting amplitude $S(\omega_N)=\sqrt{F^2+G^2}$ from the sample slice precessing at $\omega_N$.

Briefly, the signal amplitude $S(\omega_N)$ depends on the relative positions of the sample and sensor along the $y$ and $z$ directions, while $x$ is encoded in the $\omega_N$ and the $x^\prime$ dependence is dropped by the integration (Eq. \ref{eq:integral}). 

As mentioned above and described in detail in Ref.\cite{biancalana_apb_16}, we use an OAM setup in which the environmental magnetic disturbances are eliminated by applying a differential technique. The magnetometric measurement is gradiometric, and --as represented in Fig. \ref{fig:setupSM}-- the two sensors are displaced along $z^\prime$. In conclusion, $\Delta y$ is a constant, and the recorded differential signal is proportional to
\begin{equation*}
\eta= \partial S / \partial z^\prime,
\end{equation*}
which is the ideal response to a sample homogeneously magnetized along $x$, and is the coupling factor $\eta(x)$ used to reconstruct the uni-dimensional shape of the sample.

\bibliography{inhdressrefs}

\end{document}